\documentclass[fleqn,usenatbib,useAMS]{mnras}
\usepackage{threeparttablex}
\usepackage{graphicx}	
\usepackage{mwe}
\usepackage{amsmath}	
\usepackage{amssymb}	
\usepackage{multicol}   
\usepackage{bm}		
\usepackage{pdflscape}	
\usepackage{multirow}
\usepackage{ulem}
\usepackage{comment}
\usepackage{array}
\usepackage{caption}
\usepackage{subcaption}
\usepackage{lineno}
\usepackage{float}
\restylefloat{table}
\usepackage{newtxtext,newtxmath}
\includecomment{comment}

\usepackage[T1]{fontenc}
\usepackage{ae,aecompl}
\usepackage{newtxtext,newtxmath}
\usepackage{caption}
\usepackage{float}
\usepackage{hyperref}

\title[SN~2023ixf at one-year post-explosion]{Signatures of the Shock Interaction as an Additional Power Source in the Nebular Spectra of SN~2023ixf}

\author[Kumar A., et al., 2025]{\href{https://orcid.org/0000-0002-4870-9436}{Amit Kumar}$^{1,2}$\thanks{Contact: \href{mailto:amit.kumar@rhul.ac.uk}{amit.kumar@rhul.ac.uk};
\href{mailto:amitkundu515@gmail.com}{amitkundu515@gmail.com}},
\href{https://orcid.org/0000-0001-6191-7160}{Raya Dastidar}$^{3,4}$,
\href{https://orcid.org/0000-0003-0733-7215}{Justyn R. Maund}$^{1}$,
Adam J. Singleton$^{5}$ and
\href{https://orcid.org/0000-0002-4731-9698}{Ning-Chen Sun}$^{6,7,8}$
\\
$^{1}$Department of Physics, Royal Holloway - University of London, Egham, TW20 0EX, U.K. \\
$^{2}$Department of Physics, University of Warwick, Gibbet Hill Road, Coventry CV4 7AL, U.K. \\
$^{3}$Instituto de Astrof\'{i}sica, Universidad Andres Bello, Fernandez Concha 700, Las Condes, Santiago RM, Chile\\
$^{4}$Millennium Institute of Astrophysics, Nuncio Monsenor S´otero Sanz 100, Providencia, Santiago, 8320000 Chile\\
$^{5}$School of Mathematical and Physical Sciences, University of Sheffield, Sheffield, S3 7RH, U.K.\\
$^{6}$School of Astronomy and Space Science, University of Chinese Academy of Sciences, Beijing 100 049, China\\
$^{7}$National Astronomical Observatories, Chinese Academy of Sciences, Beijing 100 101, China\\
$^{8}$Institute for Frontiers in Astronomy and Astrophysics, Beijing Normal University, Beijing, 102 206, China}

\date{Accepted 2025 February 18. Received in original form 2024 December 04}

\pubyear{\the\year{}}

\begin{document}
\label{firstpage}
\pagerange{\pageref{firstpage}--\pageref{lastpage}}
\maketitle

\begin{abstract}
Red supergiants may lose significant mass during the final 100–1000 years before core collapse, shaping their circumstellar environment. The supernova (SN) shockwave propagating through this environment forms a shock-swept dense shell that interacts with the surrounding circumstellar material (CSM), generating secondary shocks that energise the ejecta and may power the SN during the nebular phase. In the present work, we investigate the nebular spectrum of SN~2023ixf, observed one-year post-explosion (at +363 d) with the recently commissioned WEAVE instrument on the 4.2m William Herschel Telescope. This marks the first supernova spectrum captured with WEAVE. In this spectrum, \ion{H}{$\alpha$} exhibits a peculiar evolution, flanked by blueward and redward broad components centred at $\sim\pm 5650\,\mathrm{km\,s^{-1}}$, features that have been observed in only a few SNe as early as one-year post-explosion. These features may indicate energy deposition from shock generated by the interaction of shock-swept dense shell with CSM expelled a few hundred years prior to the explosion. Comparisons of the +363 d spectrum with model spectra from the literature suggest a shock power of at least $\sim5 \times 10 ^{40}\,\mathrm{erg\,s^{-1}}$ at this epoch. Additionally, analysis of the [\ion{O}{i}] doublet and other emission lines helped to constrain the oxygen mass ($\lesssim 0.07-0.30 M_\odot$), He-core mass ($\lesssim 3 M_\odot$), and zero-age main sequence mass ($\lesssim 12 M_\odot$) for SN~2023ixf. The comparison with other Type II SNe highlights SN 2023ixf's unique shock interaction signatures and evidence of dust formation, setting it apart in terms of evolution and dynamics.

\end{abstract}

\begin{keywords}
transients: supernovae -- supernovae: general -- supernovae: individual: SN~2023ixf -- stars: low-mass -- techniques: spectroscopic.
\end{keywords}

\section{Introduction} \label{sec:intro}

The properties of supernovae (SNe) are primarily governed by the characteristics of their progenitors and their surrounding environments, which are closely linked. The progenitor's properties (e.g., mass, composition, metallicity, rotation, and magnetic field) not only play a critical role in determining overall ejecta morphology and the powering mechanisms---such as radioactive decay, ejecta heated by the central engine, or circumstellar material interaction (CSMI)---but also shape the surrounding environment. 

Type II SNe are thought to originate from the core-collapse of red supergiants (RSGs) with a zero-age main sequence mass ($M_{\mathrm ZAMS}$) range of $\sim 8-25 M_\odot$ \citep{Heger2003, Smartt2009, VanDyk2017}. The most direct method for probing the properties of SN progenitors is to investigate the pre-explosion images to identify the progenitor star.  This approach is only feasible, however, for nearby SNe \citep{Maund2005, Gal-Yam2007, Maund2014, Maund2017, VanDyk2017, Kilpatrick2023, Pledger2023}. For distant SNe, the comparatively lower intrinsic luminosity of progenitor stars and spatial resolution limitations make direct detection highly challenging. Upcoming advanced facilities, such as the Large Synoptic Survey Telescope (LSST; \citealt{LSST2009}) are expected to significantly improve these capabilities \citep[see][]{Strotjohann2024}. An alternative, indirect method of constraining the progenitor properties is through observations and modelling of the nebular spectra of SNe. During the nebular phase ($\gtrsim 140\,\mathrm{d}$), the spectrum is dominated by forbidden emission lines, whose luminosities, shapes and line ratios provide valuable insights into the physical conditions within the ejecta, such as temperature, density, and the mass of the emitting material. These observations can help constrain key progenitor properties such as oxygen mass ($M_{\rm O}$), which is linked to the helium-core mass ($M_{\rm He}$), $M_{\rm ZAMS}$, as well as the explosion geometry \citep{Uomoto1986, Thielemann1996, Nomoto2006, Smartt2009, Jerkstrand2014, Jerkstrand2017, Dessart2020}.

The process of mass loss significantly impacts the evolution of massive stars and also plays a crucial role in the type of the emerged SN, see \citet{Filippenko1997} and \citet{Modjaz2019} for different classes and subclasses of SNe based on the observed properties. Several potential mechanisms exist by which massive stars undergo mass loss, such as steady line-driven winds through radiation pressure \citep{Chiosi1986, Michel2023}, eruptive episodes of mass-loss \citep{Decin2006, Yoon2010, Dupree2022, Bonanos2024, Qin2024} and/or mass-transfer to a binary companion through Roche Lobe overflow \citep{Nomoto1984, Podsiadlowski1992, Drout2023, Ercolino2024}. Mass loss from the progenitor can create a dense CSM \citep{Yaron2017}, which can subsequently interact with the SN ejecta to produce enhanced luminosity and emission features with narrow cores and broad wings, depending on the properties of CSM, e.g., density, extent, distribution and mass of the CSM \citep{Schlegel1990, Stathakis1991, Kurfurst2020, Dessart2022}. These environmental effects can significantly impact the observed light curve and spectral evolution of the SN \citep{Meynet2015, Dessart2024}.

A number of SNe in the literature exhibit signatures of ejecta interacting with CSM located close to the progenitor and expelled shortly before core collapse. On the other hand, late-time spectroscopic observations of SNe that probe the mass lost by the progenitor star in the hundreds to thousands of years before explosion are rare (see \citealt{Dessart2024} for a review). At early phases, the observed properties of SNe could be mainly governed by the thermal energy in the envelope stored during the explosion, whereas other powering sources contribute at later phases, e.g., late-time ejecta-CSM interaction, decay of radioactive elements and/or centrally-located compact remanent such as magnetar \citep{Arnett1982, Chugai1990, Chevalier1994, Kasen2010, Dessart2023}. In the nebular phase spectra of some SNe, broad boxy components of \ion{H}{$\alpha$} are observed that indicate significant energy deposition by the shock generated through the interaction of the ejecta with the far-located dense CSM produced through the wind mass-loss \citep{Chugai2007, Gutierrez2017, Dessart2022, Dessart2023}. The contribution of such late-time interaction as a powering source can be seen in radio, $X-$ray and $UV$, whereas, in the optical, it can mainly be observed through \ion{H}{$\alpha$} evolution and a knee in the $R$-band light curve; see \citet{Dessart2023} and references therein. The broad boxy components of \ion{H}{$\alpha$} have been observed in the nebular spectra of a few SNe at different time-scales ($\sim$200 to 600d, e.g., SNe 1993J \citep{Matheson2000, Matheson2000a}, 2007it \citep{Andrews2011}, 2007od \citep{Andrews2010}, 2013by \citep{Black2017}, 2014G \citep{Terreran2016}, 2017eaw \citep{Weil2020}, 2017ivv \citep{Gutierrez2020}, 2020amv \citep{Sollerman2021}, and 2022jox \citep{Andrews2024}; indicating the contribution of interaction as the powering source at late-times \citep{Chugai1990, Dessart2022, Dessart2023}.

Nebular-phase spectroscopy serves as a powerful tool to probe the morphology of the ejecta, providing critical insights into asphericity through the profiles of emission lines.  Approximately half of the well-observed SNe at nebular times exhibit features of asphericity, which can also be affected by factors such as clumpiness in the ejecta or the presence of dust \citep{Maeda2008, Modjaz2008, Taubenberger2009, Mauerhan2017, Fang2024, Maund2024}.  A pristine understanding of the explosion geometry offers valuable clues about the nature of potential progenitors and the mechanisms that drive these energetic events. Forbidden emission lines, especially the [\ion{O}{i}] doublet $\lambda\lambda$6300, 6364, provide critical insights into ejecta geometry, progenitor mass, and nucleosynthesis \citep{Taubenberger2009, Jerkstrand2014, Jerkstrand2015}. In the literature \citep[e.g.,][]{Maeda2002, Mazzali2005, Maeda2008, Modjaz2008, Tanaka2009, Taubenberger2009, Fang2024}, double-peaked [\ion{O}{i}] profiles in the nebular spectra of SNe have been interpreted as signatures of oxygen-rich ejecta distributed in a toroidal or disk-like configuration, potentially viewed off-axis. However, \citet{Milisavljevic2010} highlighted challenges with this interpretation, demonstrating that many such profiles can arise from the intrinsic doublet nature of [\ion{O}{i}] $\lambda\lambda$6300, 6364, line blending, and radiative transfer effects.

The present work investigates the nebular phase spectrum of one of the closest Type II SNe, SN~2023ixf, observed nearly one-year post-explosion. SN~2023ixf was discovered on May 19, 2023 (JD 2460084.4), in the nearby galaxy M101 at a distance of approximately $6.85 \pm 0.15\,\mathrm{Mpc}$ \citep{Itagaki2023, Perley2023}. This proximity enabled extensive multi-wavelength follow-up campaigns, shedding light on the progenitor system and explosion dynamics. Initial optical and $UV$ observations captured a rapid rise in brightness driven by shock cooling and strong ejecta interaction with a dense confined CSM \citep{Bostroem2023, Hiramatsu2023, Hosseinzadeh2023, Jacobson2023, Smith2023, Teja2023, Yamanaka2023, Hu2024, Kozyreva2024, Li2024, Singh2024, Yang2024, Zimmerman2024}. Spectral flash-ionization features confirmed the presence of CSM close to the progenitor, while spectropolarimetric data revealed asymmetries in its structure \citep{Smith2023, Vasylyev2023, Li2024, Shrestha2024, Singh2024}. $X-$ray and radio observations further supported the existence of ejecta-CSM interaction at early times, providing constraints on the density and geometry of the CSM, which was expelled a few years before the core collapse \citep{Chandra2023, Grefenstette2023, Matthews2023, Iwata2024, Nayana2024, Zimmerman2024}. Pre-SN observations of the host site revealed the progenitor to be a RSG covered in dust, exhibiting periodic variability with a possible period of $\sim$1000 days, suggesting episodic mass-loss events before the explosion; however, the progenitor's initial mass remains uncertain, with estimates ranging from 8 to 24 $M_\odot$, reflecting discrepancies in analyses based on luminosity, variability, and stellar population models \citep{Jencson2023, Kilpatrick2023a, Liu2023, Niu2023, Pledger2023, Soraisam2023, Bersten2024, Fang2024b, Moriya2024, Neustadt2024, Qin2024, Ransome2024, VanDyk2024, Xiang2024}. One year of optical light curve evolution of SN~2023ixf is presented in \citet{Hsu2024}, while the latest reported nebular spectrum at +259 d post-explosion is investigated in \citet{Ferrari2024}, suggesting asymmetric ejecta and $M_{\mathrm ZAMS} \sim 12 - 15 M_\odot$.  Our current study explores the nebular spectrum at +363 d post-explosion, which represents a more developed nebular phase, providing complementary insights into the ejecta morphology and progenitor properties.

\begin{figure*}
\includegraphics[width=7cm]{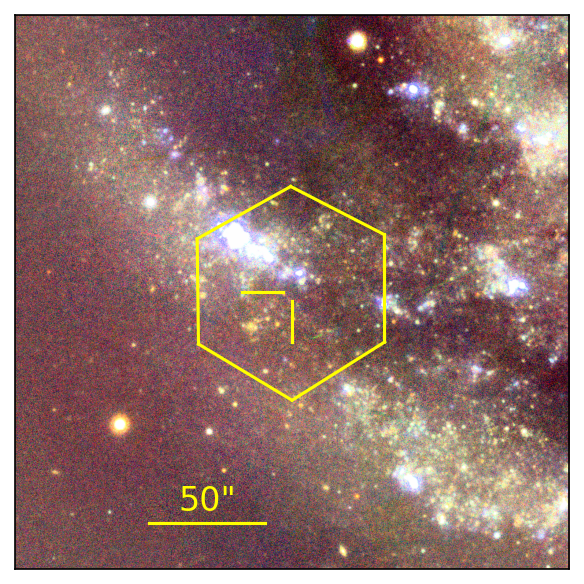}
\includegraphics[width=6.21cm]{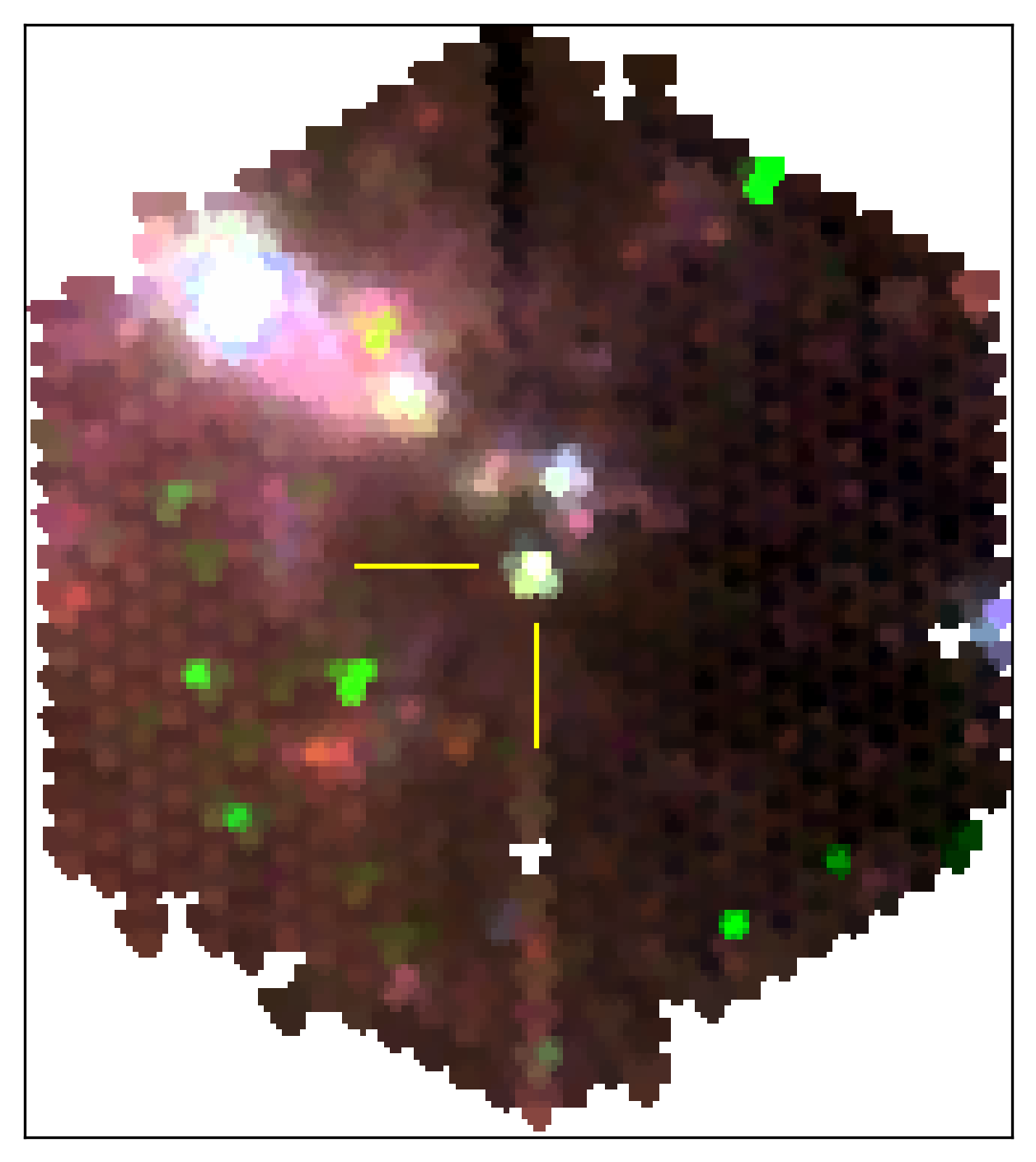}
\caption{Left panel: Pan-STARRS1 image of the site of SN~2013ixf in M101, with the footprint of the WEAVE LIFU observation indicated by the hexagon. Right panel: the WEAVE datacube, with colours generated from wavelength regions in the blue and red arms corresponding to Sloan $gri$. In both panels, the SN position is indicated by the cross-hairs.}
\label{fig:SN_site_image}
\end{figure*}

The paper is organized as follows: Section~\ref{sec:obs_data} describes the observations and data reduction; Section~\ref{sec:spec_evo} discusses the spectral evolution of SN~2023ixf from +141 to +363 days, a detailed study of the nebular lines and geometry of the ejecta; an estimate of the mass of the progenitor is presented in Section~\ref{sec:prog_mass}; shock interaction signatures are discussed in Section~\ref{sec:shock_inter}; Section~\ref{sec:spec_comp_sample} presents the comparison of the nebular spectrum (at +363 d) of SN~2023ixf with other Type II SNe at similar epochs; and Section~\ref{sec:conclusion} summarizes our findings. Throughout this study, the phase is referenced in days since the explosion.

\section{Observation and Data reduction}\label{sec:obs_data}

The site of SN~2023ixf was observed with the 4.2m William Herschel Telescope (WHT) using the WHT Enhanced Area Velocity Explorer (WEAVE) multi-object survey spectrograph \citep{Dalton2012, Dalton2016, Jin2024} on 2024 May 15 UT 23:51:16, as part of programme WS2024A2-005 (PI: Maund). WEAVE is a next-generation wide-field spectroscopic survey facility designed for high-efficiency spectral acquisition across large areas of the sky. Its Large Integral Field Unit (LIFU) mode provides spatially resolved spectroscopy over a field of view of approximately $82.9 \times 92.4~\mathrm{arcsec}^{2}$, allowing for simultaneous investigation of transient events and their host environments. WEAVE’s fibre-based design offers significant advantages in time-domain astronomy, facilitating rapid and high-quality follow-up observations. The instrument’s spectral resolution of $R\sim2500$ in the blue (3660-6060 \AA) and red (5790-9590 \AA) arms is well-suited for resolving broad and narrow spectral features in SNe, enabling precise characterization of ejecta kinematics and line profiles. Compared to other transient follow-up spectrographs, WEAVE provides a unique combination of moderate resolution and large-scale integral-field coverage, enhancing its capability to simultaneously capture the SN spectrum and map its host galaxy environment.

The observations of SN~2023ixf used WEAVE in its LIFU mode, where each fibre corresponded to a 2.6 arcsec diameter on the sky, providing spatial resolution for detailed spectral analysis. The seeing recorded at the beginning of the observation was 0.96 arcsec. The observation consisted of three dithered exposures of 1020 seconds each, utilising both the blue and red arms of WEAVE with a spectral resolution of $\approx 2500$. The combined, dithered exposures covered an area of $\approx 82.9 \times 92.4~\mathrm{arcsec}^{2}$. The final stacked datacube was retrieved from the WEAVE Operational Repository. A Pan-STARRS\footnote{https://outerspace.stsci.edu/display/PANSTARRS/Pan-STARRS1+data+archive+home+page} image of the site of SN~2023ixf in M101, overlaid with the footprint of the WEAVE LIFU observations, alongside a colour composite image derived from the WEAVE datacube itself are shown in Figure~\ref{fig:SN_site_image}.

The spectrum was extracted by summing the flux across all wavelength bins within a 3x3 pixel Region of Interest (ROI) centred on the coordinates of SN~2023ixf, using a self-developed Python script. To ensure reliable flux measurements, calibration using the sensitivity function was applied. Post-calibration, following the method discussed in \cite{Kumar2021}, the spectrum was smoothed using a Savitzky-Golay filter with a window length of 31 \AA~and a polynomial order of 3, reducing noise spikes while preserving key broad spectral features. The spectrum was shifted to the rest-wavelength assuming a redshift of M101 of z = 0.0008046 and de-reddened using $E(B-V)_{\mathrm {MW+Host}} = 0.039 \pm 0.011\,\mathrm{mag}$ derived by \citet{Singh2024}.

\begin{figure*}
\centering
	\includegraphics[width=1\textwidth]{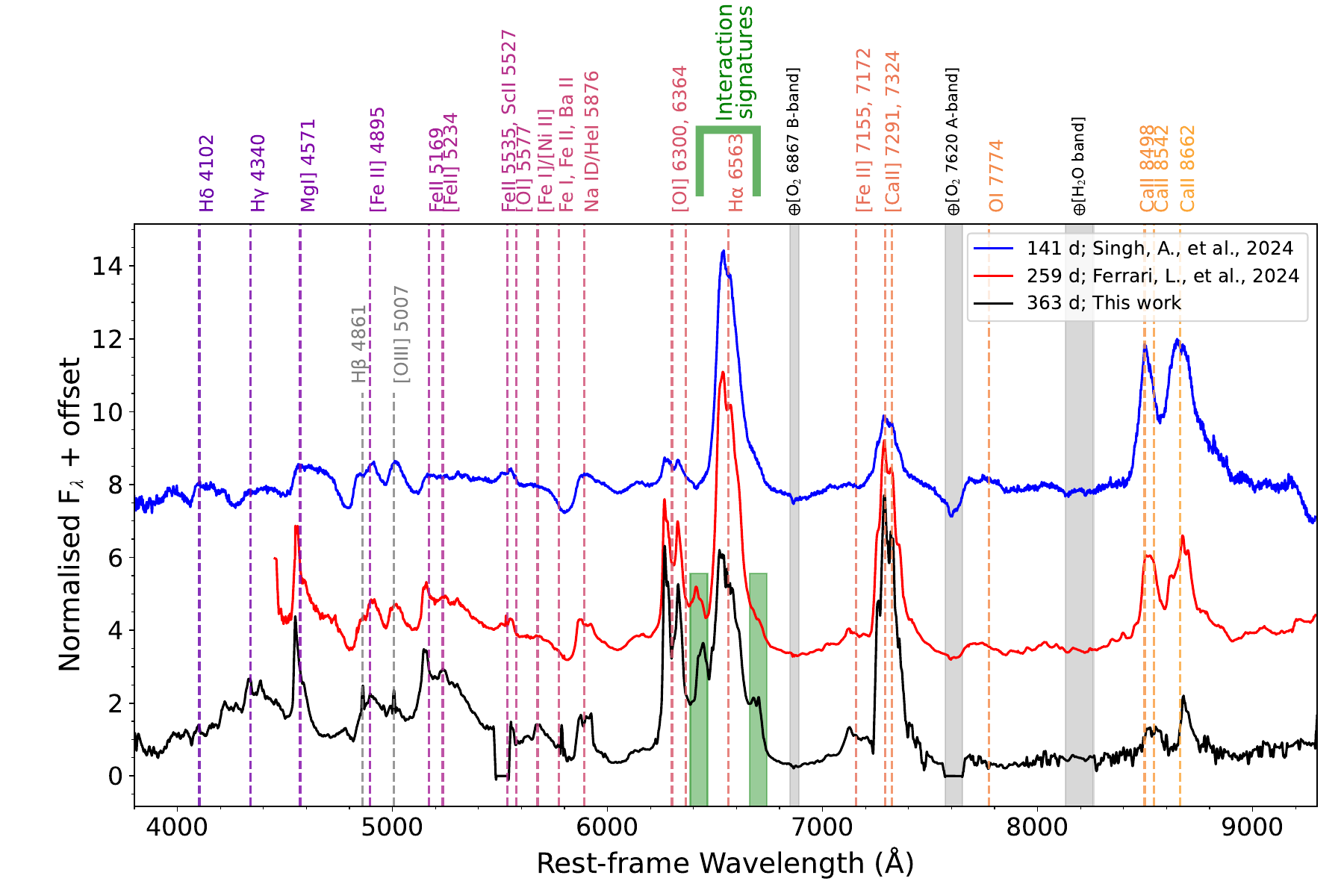}
    \caption{The comparison of the rest-frame spectra of SN~2023ixf at three different phases: +141 d \citep{Singh2024}, +259~d \citep{Ferrari2024}, and +363 d (this work). Prominent spectral features are labelled with colour-coded vertical dashed lines at their respective rest-frame wavelengths. Atmospheric absorption bands, O$_2$ and H$_2$O, are shaded in grey. Narrow emission lines corresponding to \ion{H}{$\beta$} and [\ion{O}{iii}] $\lambda$5007, identified as galaxy emission lines are highlighted with grey dashed lines. The spectra are normalised and offsets have been applied for clarity. The interaction signatures as blueward and redward components of \ion{H}{$\alpha$}, as discussed by \citet{Dessart2023}, can be seen clearly in the spectra at +259 d and +363 d, and are marked with green bands.}
\label{fig:spec_comp}
\end{figure*}

\section{Spectral evolution}\label{sec:spec_evo}

The spectral evolution of SN~2023ixf from +141 to +363 days post-explosion is presented in Figure~\ref{fig:spec_comp}. The spectra observed at +141 d and +259 d are taken from \cite{Singh2024} and \cite{Ferrari2024}, respectively. All three spectra have been normalised to their respective median flux values and a constant offset has also been applied, to facilitate clear comparisons of their features. All the prominent features are marked with colour-coded vertical lines, and their respective rest-frame wavelengths are written on the top. In the +363 d spectrum, narrow emission lines corresponding to \ion{H}{$\beta$} and [\ion{O}{iii}] $\lambda$5007 are observed at nearly zero velocity and are likely host galaxy emission lines (highlighted with grey dashed lines). Their alignment with the systemic velocity of the host supports this interpretation. However, given the ongoing evidence of shock-CSM interaction in SN~2023ixf (discussed in Section~\ref{sec:shock_inter}), these lines could also originate from the unshocked CSM. Further observations tracking the evolution of these lines are required to confirm their origin, as this study is based on a single spectrum.

Notably, the blueward component of \ion{H}{$\alpha$}, previously marked as unidentified by \cite{Ferrari2024} and not visible in the spectrum at +141 d \citep{Singh2024}, appears more prominent in our +363 d spectrum, along with a clearly visible redward component. These features are highlighted with green bands (see Figure~\ref{fig:spec_comp}), likely indicating an interaction between a shock-swept dense shell with a CSM ejected nearly 100-1000 years before the core-collapse, as suggested by \cite{Dessart2023}. A detailed analysis of these features is provided in Section~\ref{sec:shock_inter}.

Additionally, the spectrum at +363 d exhibits diminished hydrogen line intensities, particularly \ion{H}{$\alpha$}. In contrast, emission features on the blue side of 6000 \AA~become more prominent, dominated by \ion{Fe}{ii} lines, alongside increasingly visible forbidden emission lines such as \ion{Mg}{i}], the [\ion{O}{i}] doublet and [\ion{Ca}{ii}] $\lambda\lambda$7291, 7324. These features clearly indicate cooling, decreasing ejecta density, and the transition to deeper, metal-rich ejecta layers. The \ion{O}{i} $\lambda$7774 and \ion{Ca}{ii} NIR triplet show a steady decline in strength, decreasing from +141 d to +259 d and further fading at +363 d, showing decreasing ionization and recombination as the ejecta expands and cools.

\begin{figure*}
	\includegraphics[width=\textwidth]{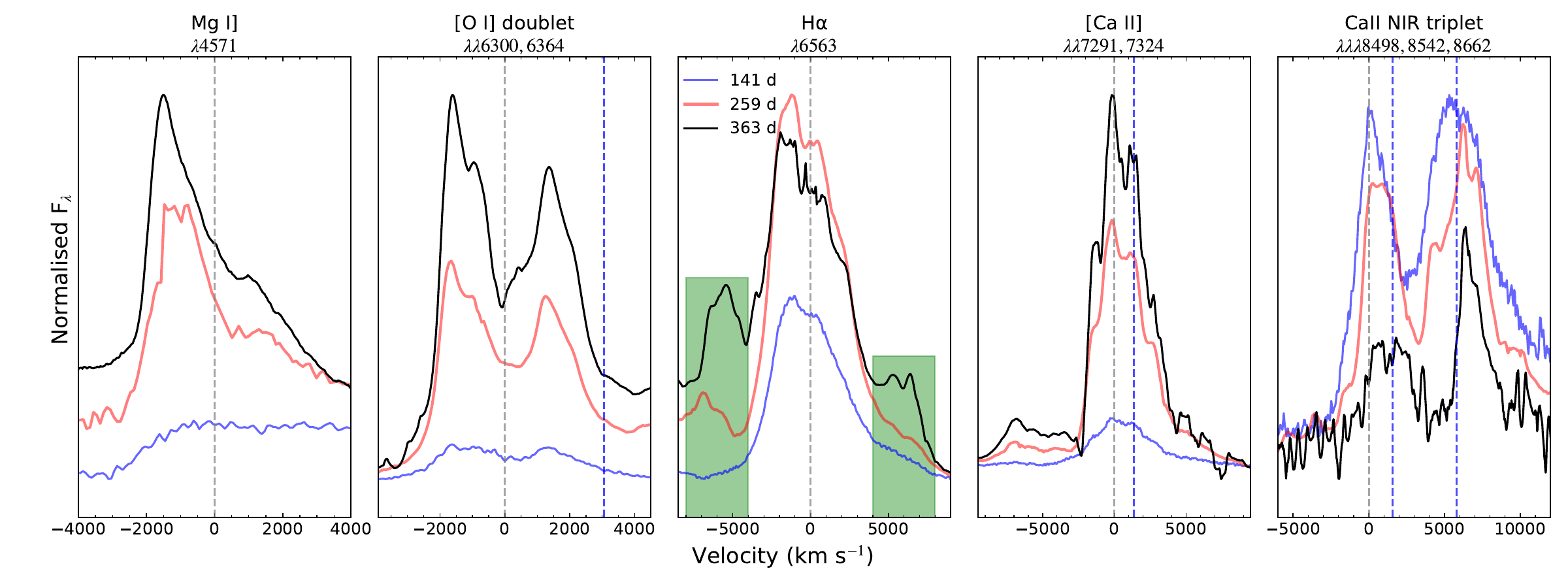}
    \caption{The evolution of \ion{Mg}{i}], [\ion{O}{i}] doublet, \ion{H}{$\alpha$}, [\ion{Ca}{ii}] $\lambda\lambda$7291, 7324, and \ion{Ca}{ii} NIR triplet in the velocity space is presented, with each line highlighted in a separate panel. The zero velocity for each line is marked at $\lambda$6300 for [\ion{O}{i}] doublet, $\lambda$6563 for \ion{H}{$\alpha$}, $\lambda$7291 for [\ion{Ca}{ii}], and $\lambda$8498 for \ion{Ca}{ii} NIR triplet in grey colour, whereas the positions of other related wavelengths are also shown with blue dashed lines in the velocity space.}
    \label{fig:line_evo_comp}
\end{figure*}

\subsection{Nebular lines study}\label{sec:Neb_lin_stu}

The five most prominent lines--\ion{Mg}{i}] $\lambda$4571, [\ion{O}{i}] doublet, \ion{H}{$\alpha$}, [\ion{Ca}{ii}] $\lambda\lambda$7291, 7324 and \ion{Ca}{ii} NIR triplet-- in the spectra at +141 d, +259 d and +363 d are compared in velocity space in Figure~\ref{fig:line_evo_comp}, providing insights into the evolution of the expansion velocities and the physical conditions of the ejecta. For \ion{Mg}{i}] $\lambda$4571, [\ion{O}{i}] doublet, \ion{H}{$\alpha$}, [\ion{Ca}{ii}] $\lambda\lambda$7291, 7324 and \ion{Ca}{ii} NIR triplet emission lines, the zero velocities are respectively taken at $\lambda$4571, $\lambda$6300, $\lambda$6563, $\lambda$7291 and $\lambda$8498 and shown with the vertical grey dashed lines. The rest wavelengths of other related lines, [\ion{O}{i}] $\lambda$6364, [\ion{Ca}{ii}] $\lambda$ 7324 and \ion{Ca}{ii} $\lambda\lambda$8542, 8662, in velocity space are marked with blue dashed lines in their respective panels. The presented features are normalised relative to the median flux values of emission-free regions adjacent to the corresponding emission lines. 

Across all three spectra (+141 d to +363 d), the [\ion{O}{i}] doublet emission features are significantly blue-shifted. In the +259 d and +363 d spectra, the \ion{Mg}{i}] $\lambda$4571 and [\ion{O}{i}] doublet are shifted by approximately $-1500\,\mathrm{km\,s^{-1}}$ from their respective rest wavelengths. The \ion{H}{$\alpha$} line exhibits multiple components, including a central broad profile surrounded by boxy blueward and redward features (marked in green; see the third panel of Figure~\ref{fig:line_evo_comp}), each displaying a multi-peak structure in the +363 d spectrum. From +141 d to +363 d, the central \ion{H}{$\alpha$} profile is observed nearly at zero velocity, though it appears asymmetric around the central wavelength. In the spectrum at +259 d, the blueward component of \ion{H}{$\alpha$} line is observed at a velocity around $-6900\,\mathrm{km\,s^{-1}}$, while the redward component is not visible. By + 363 d, this blueward component of \ion{H}{$\alpha$} shifts closer to the rest velocity of \ion{H}{$\alpha$}, peaking at $\sim-5650\,\mathrm{km\,s^{-1}}$. Meanwhile, the redward component emerges at nearly the same velocity but in the opposite direction ($\sim +5650\,\mathrm{km\,s^{-1}}$), albeit comparatively suppressed. The asymmetry of the central \ion{H}{$\alpha$} profile around zero velocity, along with the distinct profiles of the blueward and redward components, indicate asymmetry and clumpiness in the ejecta and/or surrounding CSM; discussed in detail in Section~\ref{sec:shock_inter}. 

The [\ion{Ca}{ii}] $\lambda\lambda$7291, 7324 features appear slightly blue-shifted, with velocities in the range of approximately $-100$ to $-170\,\mathrm{km\,s^{-1}}$, and increase in intensity over time. Additionally, the emergence of [\ion{Fe}{ii}] $\lambda\lambda$7155, 7172 on the left of this feature becomes apparent. On the other hand, the \ion{Ca}{ii} NIR triplet appears nearly at zero velocity in the +141 d spectrum but becomes redshifted in the +259 d and +363 d spectra ($\approx 600\,\mathrm{km\, s^{-1}}$), while simultaneously weakening over time.

\begin{figure}
	\includegraphics[width=0.48\textwidth]{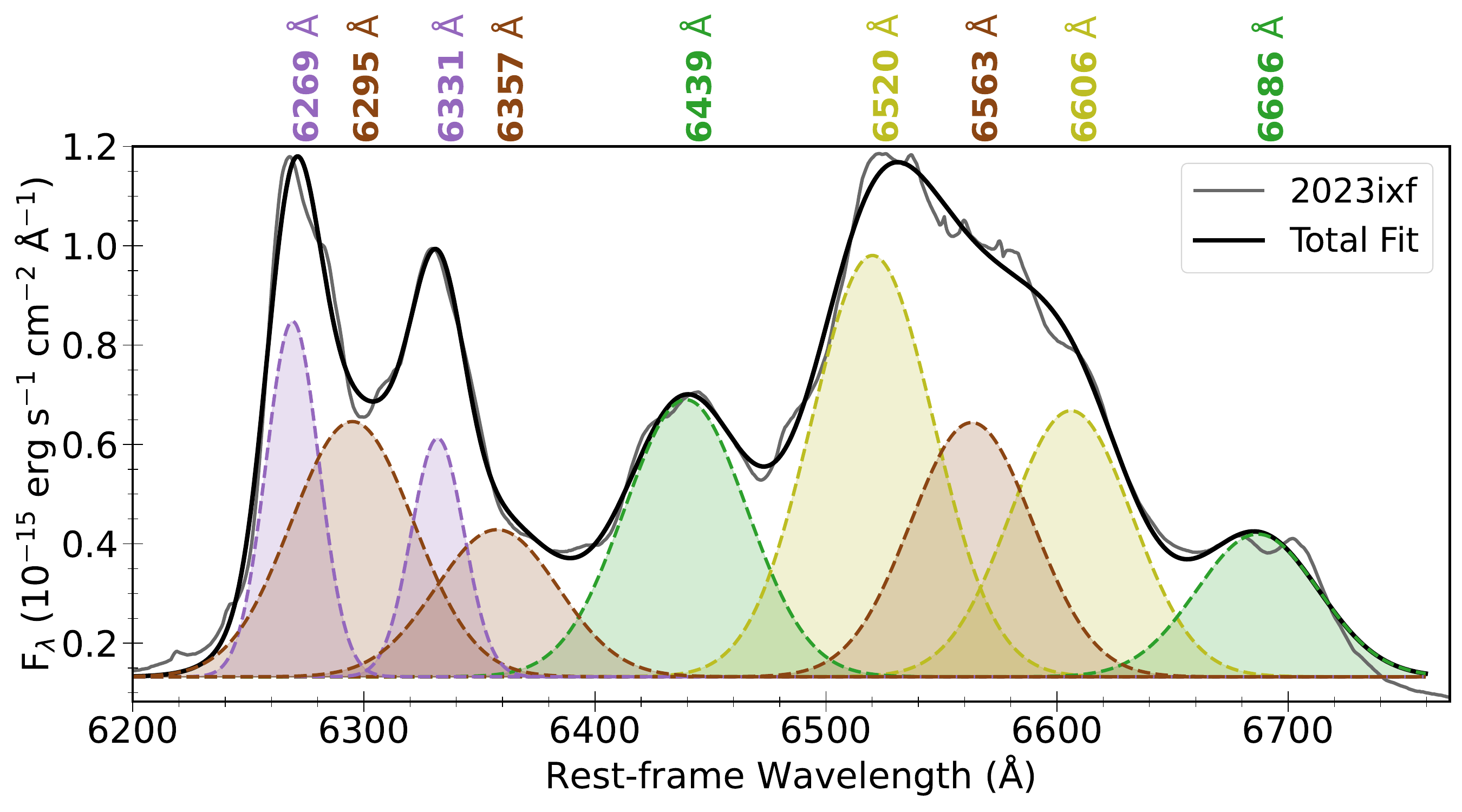}
    \caption{Multiple Gaussian components fitted to the [\ion{O}{i}] doublet and \ion{H}{$\alpha$} profile are shown. The Gaussian components near the rest wavelengths of the [\ion{O}{i}] doublet and \ion{H}{$\alpha$} are shaded in brown. The central wavelengths of each fitted Gaussian are labelled at the top.}
    \label{fig:OI_fit}
\end{figure}

To analyse the blended [\ion{O}{i}] doublet and \ion{H}{$\alpha$} composite structure, we perform a simultaneous multi-Gaussian fit in the wavelength region 6200 to 6750 \AA~using the Python package \texttt{scipy.curve\_fit} (see Figure~\ref{fig:OI_fit}). For the \ion{H}{$\alpha$} line, we fit the central component with a Gaussian initialised at 6562.8 \AA. Additionally, we include two sets of satellite Gaussians symmetrically shifted by variable parameters, $\Delta\lambda_{01}$ and $\Delta\lambda_{02}$, on both sides of the central \ion{H}{$\alpha$} component. These shift parameters ($\Delta\lambda_{01}$ and $\Delta\lambda_{02}$) are treated as free variables during the fitting process. All Gaussians associated with \ion{H}{$\alpha$} are constrained to have the same full width at half maximum (FWHM). For the [\ion{O}{i}] doublet, each component is fitted using two Gaussians: a narrow Gaussian and a broad Gaussian. The broad Gaussian is constrained to have the same FWHM as the \ion{H}{$\alpha$} Gaussians, while the FWHM of the narrow Gaussian is treated as an independent free parameter. The individual Gaussian components and the total fit are shown in Figure~\ref{fig:OI_fit}, with the central wavelengths of the fitted Gaussians marked above the top. The parameter estimates from the multi-Gaussian fitting process are provided in Table~\ref{tab:GaussianFitResults}. From this fit, we determined the FWHM values for the broad and narrow components to be $62.3 \pm 2.7$ \AA~and $26.9 \pm 2.6$ \AA, respectively. The shift parameters $\Delta\lambda_{01}$ and $\Delta\lambda_{02}$ were found to be $\pm (42.9\pm2.6)$ \AA~and $\pm (123.7 \pm 2.6)$~\AA, respectively. We note that the bridge-like feature connecting [\ion{O}{i}] doublet and \ion{H}{$\alpha$} has contributions from both the blueward component \ion{H}{$\alpha$} and the redward component of [\ion{O}{i}] doublet. These features indicate a complex velocity structure of the material within the ejecta, with components moving at varying speeds \citep{Kuncarayakti2020}. From these fits, we measured the flux of the [\ion{O}{i}] doublet, calculated as the composite flux of all four Gaussian components, to be $0.93 \times 10^{-13}\,\mathrm{erg\,cm^{-2}\,s^{-1}}$, which is later used in Section~\ref{sec:oi_doublet}.

\begin{table*}
\centering
\caption{Parameter estimates of the multi-Gaussian fitting process for the blended [\ion{O}{i}] doublet and \ion{H}{$\alpha$} composite structure, and \ion{Mg}{i}] profile.}
\label{tab:GaussianFitResults}
\begin{tabular}{|c|l|l|p{6cm}|}
\hline
\textbf{Component}            & \textbf{Parameter}         & \textbf{Best-fitted value (\AA)}  & \textbf{Notes} \\ \hline
\multirow{4}{*}{\ion{H}{$\alpha$}} 
                              & Central wavelength         & 6563.1 $\pm$ 2.6                & Central Gaussian component of \ion{H}{$\alpha$}. \\ \cline{2-4} 
                              & FWHM                      & 62.3 $\pm$ 2.7            & Broad component, same FWHM applied to all \ion{H}{$\alpha$} Gaussians. \\ \cline{2-4} 
                              & $\Delta\lambda_{01}$       & ± (42.9 $\pm$ 2.6)            & Shift for the first pair of satellite components. \\ \cline{2-4} 
                              & $\Delta\lambda_{02}$       & ± (123.7 $\pm$ 2.6)            & Shift for the second pair of satellite components. \\ \hline
\multirow{3}{*}{[\ion{O}{i}] Doublet} 
                              & Central wavelengths (Broad Gaussians) & 6295.0 $\pm$ 2.8, 6357.6 $\pm$ 2.9 & \multirow{1}{*}{Broad component constrained to same FWHM} \\ \cline{2-3}
                              & FWHM (Broad Gaussians)      & 62.3 $\pm$ 2.6            & \multirow{1}{*}{as \ion{H}{$\alpha$}}. \\ \cline{2-4} 
                              & Central wavelengths (Narrow Gaussians) & 6269.1 $\pm$ 2.5, 6331.9 $\pm$ 2.5 & Independent narrow components. \\ \cline{2-3}
                              & FWHM (Narrow Gaussians)     & 26.9 $\pm$ 2.6            &  \\ \hline
\multirow{3}{*}{\ion{Mg}{i}]} 
                              & Central wavelength (Broad Gaussian) & 4578.5 $\pm$ 2.0 & \multirow{1}{*}{Broad component.} \\ \cline{2-3}
                              & FWHM (Broad Gaussian)      & 57.5 $\pm$ 2.2            &  \\ \cline{2-4} 
                              & Central wavelength (Narrow Gaussian) & 4550.4 $\pm$ 1.8 & Narrow component. \\ \cline{2-3}
                              & FWHM (Narrow Gaussian)     & 20.0 $\pm$ 1.9            &  \\ \hline
\end{tabular}
\end{table*}

\subsubsection{Geometry}

Profiles of emission lines in the nebular spectra of SNe are powerful tools for probing the geometry of the ejecta. Among these, the [\ion{O}{i}] doublet $\lambda\lambda$6300, 6364 is particularly useful due to its high strength during the nebular phase and its relative isolation from blending with other lines. Based on numerous studies \citep[e.g.,][]{Maeda2002, Mazzali2005, Maeda2008, Modjaz2008, Tanaka2009, Taubenberger2009, Fang2024}, double-peaked [\ion{O}{i}] profiles have often been interpreted as arising from oxygen-rich ejecta arranged in toroidal or disk-like geometry, potentially viewed off-axis. However, \citet{Milisavljevic2010} highlighted the challenges of interpreting [\ion{O}{i}] doublet profiles solely as evidence of such a geometry, emphasizing that the separation of the double peaks in the [\ion{O}{i}] doublet plays a crucial role in the interpretation.

For SN~2023ixf, the spectrum at +359 d (Figures~\ref{fig:line_evo_comp} and \ref{fig:OI_fit}) shows that the narrow components of the [\ion{O}{i}] doublet are separated by $\approx$63 $\pm$ 1 \AA, with both peaks equidistant from 6300 \AA~(the zero-velocity position). This separation matches the theoretically expected value for the [\ion{O}{i}] $\lambda$6300 and [\ion{O}{i}] $\lambda$6364 lines. In this context, the intrinsic doublet nature of the [\ion{O}{i}] lines is sufficient to explain the symmetric double-peaked profiles without invoking a toroidal or disk-like geometry. Furthermore, from the fit, it is apparent that the narrow components of the [\ion{O}{i}] doublet, shifted by $-$1500 km s$^{-1}$, contribute to the overall blue-shifted structure, while the peaks of the broad components remain near their rest wavelengths. The blue-shifted, narrow profiles and the broad, nearly zero velocity profiles could potentially originate from two different regions of the SN: a central, spherically symmetric distribution of O-rich ejecta and a clump or shell of O-rich material moving at a moderate velocity ($-$1500 km s$^{-1}$) toward the observer.

Additionally, the intensity ratio between the [\ion{O}{i}] $\lambda$6300 and [\ion{O}{i}] $\lambda$6364 components is measured to be $\approx$1.15, significantly lower than the theoretical value of 3 for an optically thin region. This reduced ratio suggests suppression of the bluer $\lambda$6300 component, likely due to optically thick line emission with $\tau > 1$ \citep{Li1992, Williams1994}. Further evidence supporting the identification of the two peaks as the [\ion{O}{i}] $\lambda\lambda$6300, 6364 doublet comes from the increasing blue to red peak height ratio from +141 d to +363 d, as the SN's expansion (and corresponding decrease in opacity) should drive this ratio toward the expected nebular-phase 3:1 ratio.

Interestingly, we also noted that the peak of \ion{Mg}{i}] is also blue-shifted by the same amount as [\ion{O}{i}]. Hydrodynamic explosion models \citep{Maeda2006} predict that Mg and O should share a similar spatial distribution within the SN ejecta, while heavier elements such as Fe and Ca may exhibit significantly different distributions. As a result, the profiles of isolated Mg and O emission lines are expected to be similar, a trend observed in several stripped-envelope SNe \citep{Taubenberger2009}.

However, a direct comparison of the \ion{Mg}{i}] and [\ion{O}{i}] lines is challenging because, unlike \ion{Mg}{i}], the [\ion{O}{i}] feature is a doublet. To address this, \citet{Taubenberger2009} proposed a method in which the \ion{Mg}{i}] feature is first isolated, and a linearly fitted background is subtracted. The \ion{Mg}{i}] line is then rescaled to one-third of its initial intensity, shifted by 46 \AA~(equivalent to the 64 \AA~separation between the two [\ion{O}{i}] lines), and added back to the original profile. This modified \ion{Mg}{i}] profile can then be directly compared with the observed [\ion{O}{i}] feature.

\begin{figure}
	\includegraphics[width=0.48\textwidth]{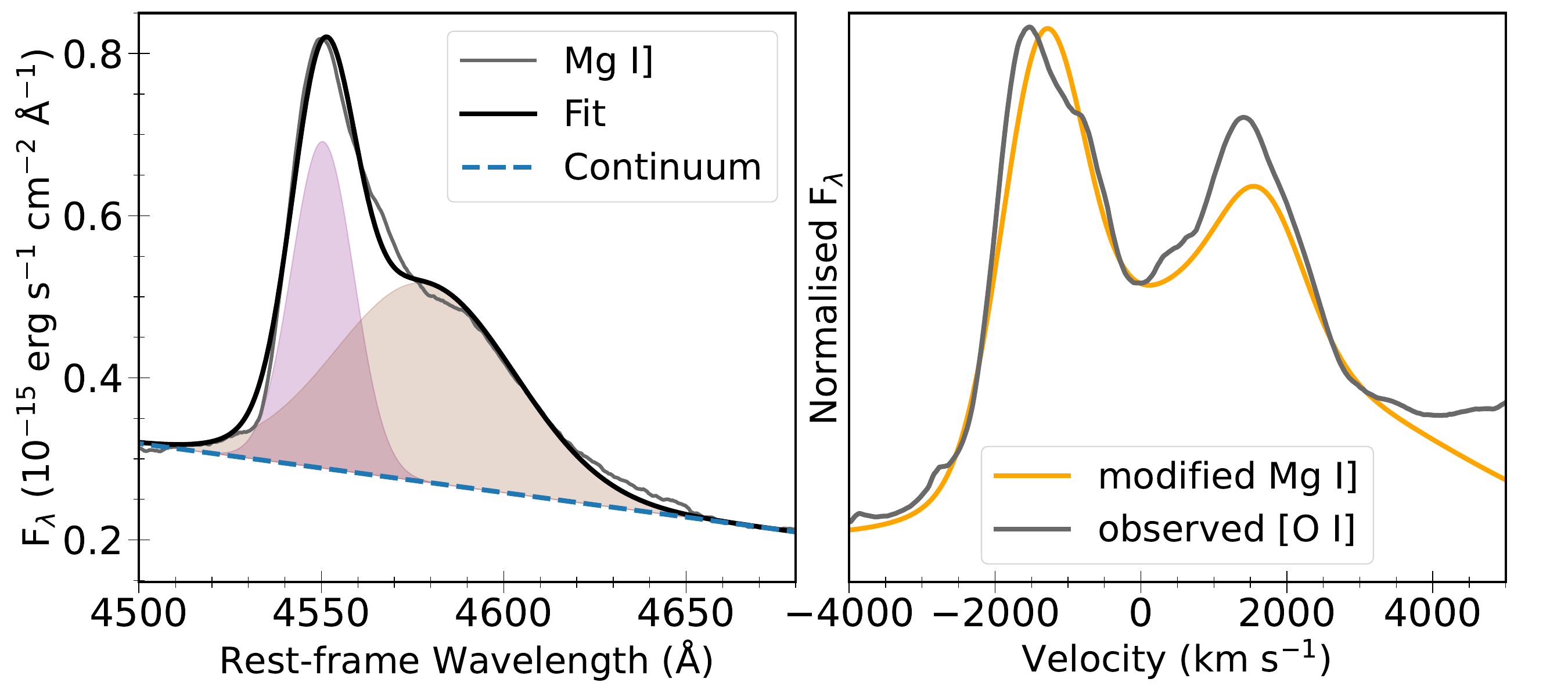}
    \caption{Double Gaussian fitted to the \ion{Mg}{i}] doublet is shown along with the linearly fitted continuum. The modified \ion{Mg}{i}] profile is compared in the right panel with the observed [\ion{O}{i}] profile.}
    \label{fig:MgI_fit}
\end{figure}

For SN~2023ixf, we initially fit the \ion{Mg}{i}] profile using both a narrow and a broad Gaussian component, as shown in the left panel of Figure~\ref{fig:MgI_fit}. We applied the rescaling to both components, and the final modified profile is presented in the right panel of Figure~\ref{fig:MgI_fit}. The modified \ion{Mg}{i}] profile closely reproduces the observed [\ion{O}{i}] feature, except that the $\lambda$6364 component of [\ion{O}{i}] appears brighter than its counterpart in the modified \ion{Mg}{i}] profile. This discrepancy is expected given the lower-than-expected [\ion{O}{i}] doublet ratio ($\sim$1.15) obtained for SN~2023ixf. Thus, the similarity between the [\ion{O}{i}] and \ion{Mg}{i}] profiles suggests a similar spatial distribution of these elements and further supports the presence of a directional component in addition to a centrally symmetric ejecta structure. The latter is powered by radioactive decay and also contributes to the central \ion{H}{$\alpha$} profile centred at 6563 \AA~(discussed in more detail in Section~\ref{sec:shock_inter}).

\section{Progenitor mass}\label{sec:prog_mass}

The values for $M_{\rm O}$, $M_{\rm He}$, and $M_{\rm ZAMS}$ can be estimated using the observed flux of the [\ion{O}{i}] doublet, in conjunction with a comparison to the modelled spectra of \citet{Jerkstrand2014}.

\subsection{Progenitor mass from [O~{\sc i}] doublet flux}\label{sec:oi_doublet}

Forbidden oxygen lines in the nebular phase of SNe, mainly the isolated [\ion{O}{i}] doublet $\lambda\lambda$6300, 6364, are directly linked to nucleosynthesis and are considered a reliable indicator of the progenitor mass. These lines also help constrain various ejecta properties, such as mixing, the mass of synthesized $^{\rm 56}$Ni, and geometry \citep{Taubenberger2009, Jerkstrand2014, Jerkstrand2017}. \citet{Ferrari2024} have measured $M_{\rm O}$ for SN~2023ixf using the spectrum at +259 d. In this work, we also attempt to estimate $M_{\rm O}$ using our +363 d spectrum, which appears to represent a more developed nebular phase, complementing the earlier measurement. Following \citet{Uomoto1986}, $M_{\rm O}$ can be expressed as:

\begin{equation}\label{eq:oxygenmass}
M_{\rm O} = f\left([{\rm O}~{\rm I}]\right) \times d^2 \times \exp \left( \frac{2.28}{T_4} \right) \times 10^8
\end{equation}

where, $f\left([{\rm O}~{\rm I}]\right)$ is the integrated flux of the [\ion{O}{i}] doublet, $d$ is the distance to the host galaxy, and $T_4$ is the temperature of the oxygen-emitting region in $10^4$ K. Before the flux measurements, the spectrum has been corrected for the Milky Way and the host galaxy extinctions, considering $E(B-V)_{\rm MW+Host}$ = 0.039 $\pm$ 0.011 mag \citep{Singh2024}. We estimated $f\left([{\rm O}~{\rm I}]\right)$ to be $\approx0.93 \times 10^{-13} \, \mathrm{erg \, cm^{-2} \, s^{-1}}$. Similar to \citet{Singh2024}, we adopted the distance of the host galaxy of SN~2023ixf equals to 6.82 $\pm$ 0.14 Mpc, which is the mean distance of those measured by \cite{Tikhonov2015} and \cite{Riess2022}. $T_4$ can be estimated using the flux ratio of [\ion{O}{i}] $\lambda$5577 to the [\ion{O}{i}] doublet ($\lambda\lambda$6300, 6364), which is mainly sensitive to temperature and optical depth. If we assume the region is optically thin, the ratio depends solely on the temperature of the emitting region \citep{Elmhamdi2011, Jerkstrand2014}. However, in our +363 d spectrum, the region around [\ion{O}{i}] $\lambda$5577 falls within the chip gap, resulting in distortion and preventing a reliable flux measurement. Consequently, we adopt the value of $T$ with a range of 3500–4500 K (as suggested by \citealt{Bose2013}), consistent with the typical oxygen-emitting region temperatures observed in Type II SNe during the nebular phase (300–400 days; see \citealt{Liu1995, Smartt2009b, Bose2013, Jerkstrand2014, Jerkstrand2015}). This range reflects the balance of radioactive heating and line cooling processes, which are largely uniform across Type IIP SNe with similar progenitor types. Adopting this range ensures systematic uncertainties are considered while leveraging consistent thermal evolution trends observed in similar SNe. Using Equation~\ref{eq:oxygenmass} and parameter values, we estimate $M_{\rm O} \approx 0.07-0.30\,M_\odot$. Based on \citet{Jerkstrand2015}, $M_{\rm O}$ values of $0.3 - 1.3 M_\odot$ correspond to $M_{\rm ZAMS}$ of $12 - 17 M_\odot$ and $M_{\rm He}$ of $3 - 5\,M_\odot$, respectively. Based on this, SN~2023ixf likely originates from a progenitor with $M_{\rm ZAMS} \lesssim 12 M_\odot$ and $M_{\rm He} \lesssim 3 \,M_\odot$.

\begin{figure}
	\includegraphics[width=0.5\textwidth]{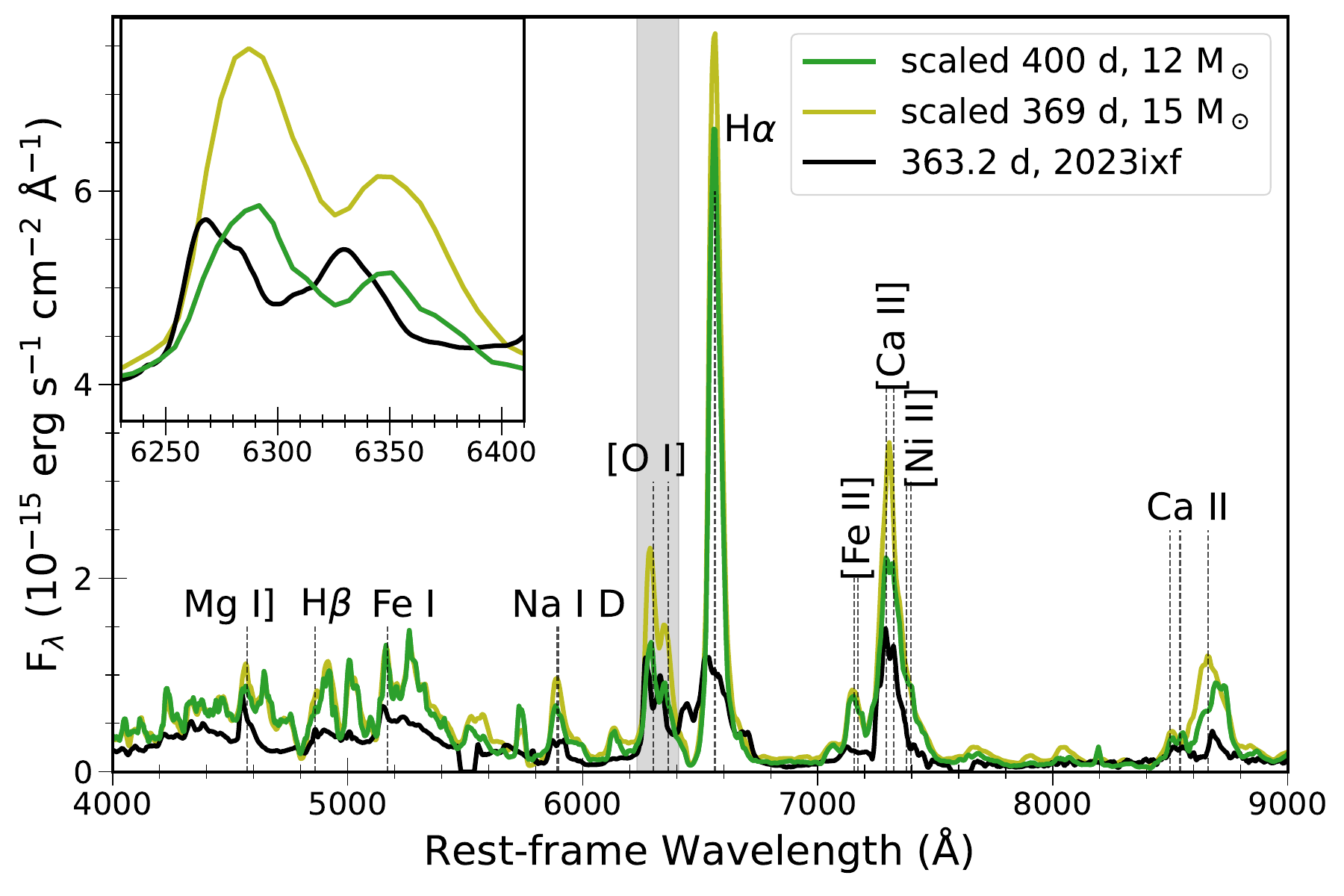}   
    \caption{Comparison of +363 d spectrum of SN~2023ixf with 12 and 15 $M_\odot$ model spectra from \citet{Jerkstrand2014}. The inset shows the zoom-in of the [\ion{O}{i}] $\lambda\lambda$6300, 6364 region.}
    \label{fig:spec_comp_jerkstrand}
\end{figure}

\subsection{Progenitor mass from literature model spectra comparison}

\citet{Jerkstrand2014} modelled nebular phase spectra of Type II SNe for progenitors with initial masses of 12, 15, 19, and 25 $M_\odot$, assuming a $^{\rm 56}$Ni mass of 0.062 $M_\odot$ and a distance of 5.5 Mpc. We compared the +363 d spectrum of SN~2023ixf to the 12 and 15 $M_\odot$ models to estimate the progenitor's mass, as shown in Figure~\ref{fig:spec_comp_jerkstrand}. During the nebular phase, the SN is powered by radioactive decay energy. To account for differences in $^{\rm 56}$Ni mass, we scaled the model spectra flux by the ratio of the $^{\rm 56}$Ni mass of SN~2023ixf (0.070 $M_\odot$, \citealt{Zimmerman2024, Hsu2024}) to that of the model spectra (0.062 $M_\odot$). The model flux was also rescaled to match the distance (6.82 Mpc) and phase (+363 d) of the nebular spectrum of SN~2023ixf. Before comparison, we corrected the SN spectrum for redshift, scaled it to match the photometric flux to account for slit losses, and applied reddening corrections ($A_V = 0.12\,\mathrm{mag}$). The model spectra display strong \ion{H}{$\alpha$} emission, which is significantly weaker in the observed spectrum of SN~2023ixf. The \ion{Mg}{i}] $\lambda$4571 line is reproduced well by both the models, but the models exhibit a prominent \ion{Ca}{II} NIR triplet feature that is weaker in the SN~2023ixf spectrum. The [\ion{O}{i}] doublet flux in the SN~2023ixf spectrum closely matches the flux predicted by the 12 $M_\odot$ model spectrum, suggesting a $M_{\rm ZAMS}$ of $\sim$12 $M_\odot$ for SN~2023ixf. Additionally, the [\ion{Ca}{ii}] in the 12 $M_\odot$ model spectrum is a better match to the observed spectrum, although the strength of this line is only weakly dependent on the progenitor mass \citep{Jerkstrand2014}.

In summary, our estimate of the progenitor mass for SN~2023ixf, derived from the flux measurement of the [\ion{O}{i}] doublet and constrained through comparison with model spectra, suggests a $M_{\text{ZAMS}}$ of $\lesssim12\, M_\odot$. This is consistent with previous studies that suggested a low-mass progenitor through pre-explosion imaging and the hydrodynamical modelling, such as \citet[][$11\pm2 \,M_\odot$]{Kilpatrick2023}, \citet[][$\sim12 \, M_\odot$]{Pledger2023}, \citet[][$9.3 - 13.6 \, M_{\odot}$]{Neustadt2024}, \citet[][$\sim10\, M_\odot$]{Singh2024} and \citet[][$12^{+2}_{-1}\,M_\odot$]{Xiang2024}.

\begin{figure}
    \includegraphics[width=0.5\textwidth]{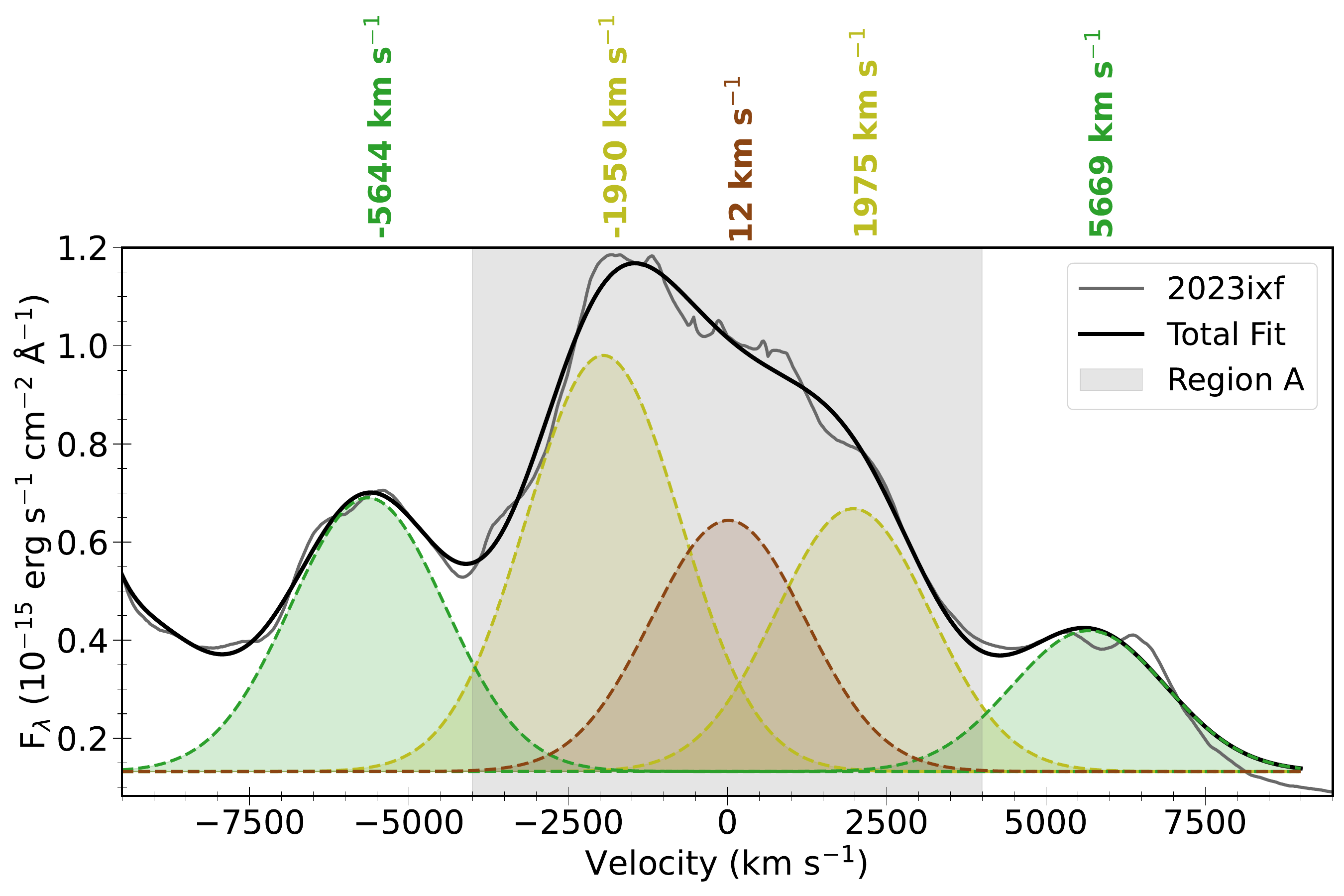}
    \caption{Zoom-in of the \ion{H}{$\alpha$} profile fit in velocity space. The velocity of each of the Gaussian components relative to the rest velocity of \ion{H}{$\alpha$} $\lambda$6562.8 are labelled at the top. The central complex \ion{H}{$\alpha$} profile, referred to as `Region A' in the text, is shaded in grey.}
    \label{fig:Ha_fit}
\end{figure}

\begin{figure*}
    \centering
	\includegraphics[width=1\textwidth]{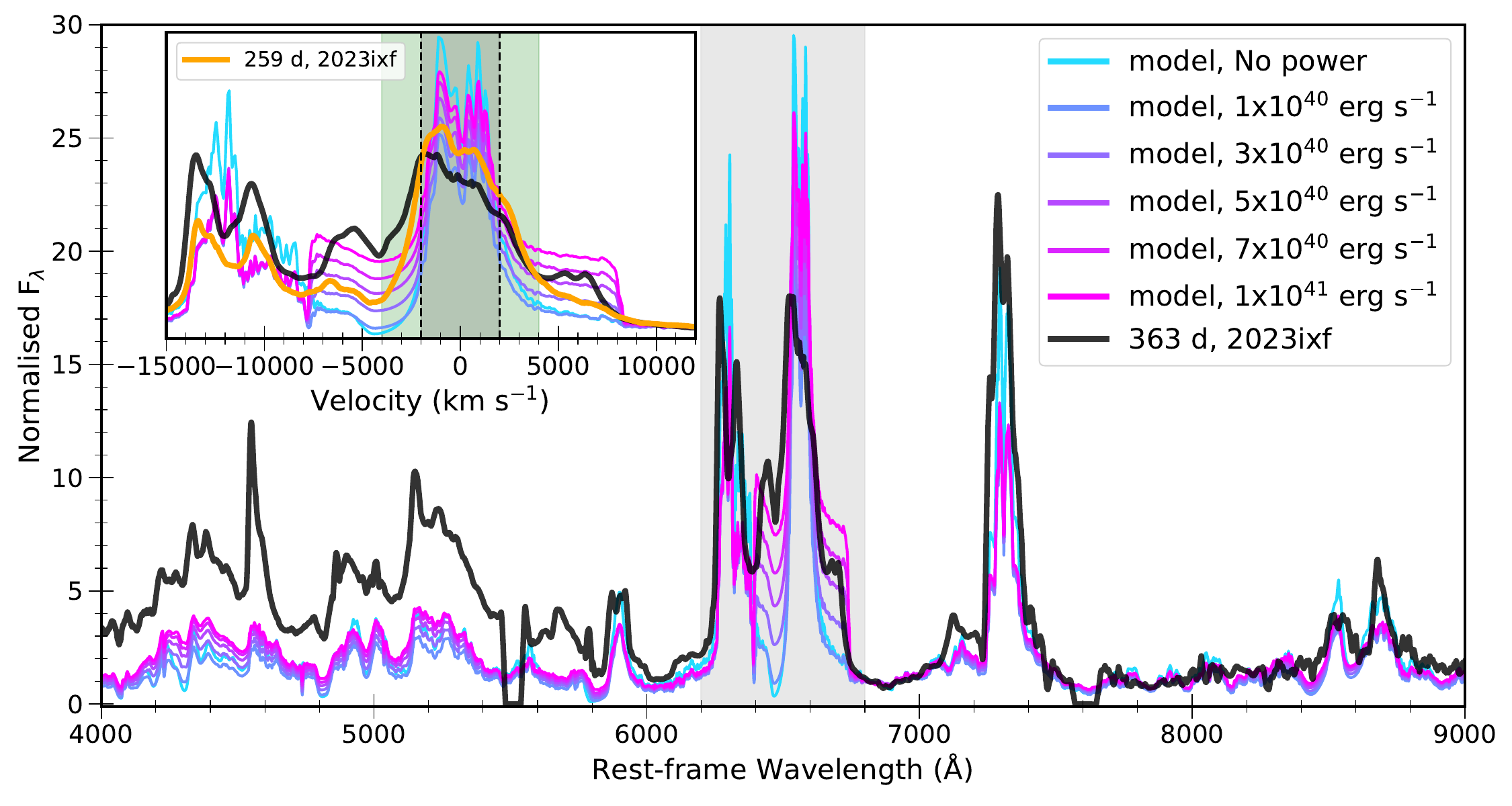}
    \caption{Comparison of the +363 d spectrum of SN~2023ixf with the +350 d nebular spectra models from \citet{Dessart2023}. The models, representing various injected shock powers ranging from $10^{40}$ to $10^{41}\,\mathrm{erg\,s^{-1}}$, are indicated in the figure legend. The inset provides a close-up view of the [\ion{O}{i}] doublet and \ion{H}{$\alpha$} regions. For comparison, the +259 d spectrum of SN~2023ixf is also shown in the inset. The extent of the central \ion{H}{$\alpha$} component in the models ($\pm2000\,\mathrm{km\, s^{-1}}$) is marked with dashed lines, while the green shaded region highlights its extent in SN~2023ixf.}
    \label{fig:Comp_Dessart}
\end{figure*}

\section{Interaction signatures}\label{sec:shock_inter}

The +363 d spectrum of SN~2023ixf exhibits a highly complex \ion{H}{$\alpha$} profile, revealing crucial details about the interaction processes within the SN environment. The central component of this structure, referred to hereafter as `Region A', can be decomposed into three distinct components, as shown in  Figure~\ref{fig:Ha_fit}, spanning velocities from $-$4000 to $+4000\,\mathrm{km\,s^{-1}}$ relative to the rest velocity of \ion{H}{$\alpha$}. The central component of `Region A', redshifted by $12\,\mathrm{km\,s^{-1}}$, is flanked by two components centred at $\sim \pm1950\,\mathrm{km\,s^{-1}}$. The blue-shifted component is notably stronger than its redshifted counterpart. A similar multi-peaked profile is also evident in the +141 d and +259 d spectra (see the third panel of Figure~\ref{fig:line_evo_comp}). In addition to `Region A', distinct high-velocity features emerge, symmetrically positioned on both sides of `Region A', centred at $\sim \pm 5650\, \mathrm{km\,s^{-1}}$ relative to the rest velocity of \ion{H}{$\alpha$}. These high-velocity features are absent in spectra obtained up to +141 d (see the third panel of Figure~\ref{fig:line_evo_comp}). The high-velocity feature on the blue side becomes discernible as early as +259 d, while that on the red side becomes prominent at +363 d. 

\cite{Dessart2023} modelled nebular-phase spectra ($>$ +350 d) of a $15.2\, M_\odot$ RSG star explosion, incorporating shock power generated by the interaction of a shock-swept dense shell with a tenuous CSM, in addition to energy from radioactive decay. To mimic this interaction, they inject shock power in the range $10^{40}-10^{41}\,\mathrm{erg \,s^{-1}}$ in a dense shell expanding at a velocity of $8000\,\mathrm{km\,s^{-1}}$. These shock powers correspond to a mass-loss rate of the progenitor in the range $10^{-6}-10^{-5}\, M_\odot\,\mathrm{yr}^{-1}$ during hundreds to thousands of years before the explosion. This injected power is comparable to the energy produced by radioactive decay one-year post-explosion, and as the radioactive energy declines over time, the shock interaction increasingly dominates the luminosity. The signatures of this interaction are evident across various wavelengths, with the $UV$ being particularly sensitive. In the $UV$, the interaction manifests as an increase in the overall continuum luminosity and enhanced emission from the \ion{Mg}{ii} $\lambda\lambda$ 2795, 2802 doublet. In the optical spectrum, the shock interaction signatures are observed as blue-shifted and red-shifted broad, boxy \ion{H}{$\alpha$} emission features.

In Figure~\ref{fig:Comp_Dessart}, we compare the +363 d spectrum of SN~2023ixf with the +350 d model spectra of \cite{Dessart2023}, which incorporate various injected shock powers ranging from $10^{40}$ to $10^{41}\,\mathrm{erg\, s^{-1}}$ in a spherical dense shell. The models exhibit a central \ion{H}{$\alpha$} profile spanning $\pm 2000\,\mathrm{km\, s^{-1}}$ (indicated by dashed lines in the inset plot), surrounded by broad, boxy emission profiles on both sides, with steep cutoffs at $\pm 8000\,\mathrm{km\, s^{-1}}$. In their models, the boxy profiles originate exclusively from the narrow dense shell and display sharp vertical edges due to the high density and limited width of the emitting region.

In SN~2023ixf, however, the central component is clearly broader and, as discussed above, blends with the boxy structure at $\pm 4000\,\mathrm{km\,s^{-1}}$. Such a multi-component profile could originate from an asymmetric, clumpy shell interacting with a low-density CSM. \cite{Singh2024} proposed the presence of an extended CSM in addition to the torus-shaped dense CSM responsible for early flash-ionisation features. According to their model, varying densities along the poles and equator result in an aspherical shock front after it breaks out of the dense CSM approximately 4 days post-explosion.

Multiple pieces of evidence for this aspherical shock front are presented in \cite{Singh2024}. After breaking out of the torus-shaped dense CSM, they describe a faster shock front moving at 13500 km s$^{-1}$, resulting from the freely expanding ejecta interacting with the low-density extended CSM, and a slower shock front, decelerated by the equatorial dense CSM, moving at 8500 km s$^{-1}$. Over time, the asymmetric shock front sweeps up material, forming an asymmetric dense shell. Both shock fronts will decelerate over time, with the slower-moving shock front accumulating more material than the faster one, leading to enhanced interaction power at lower velocities.

The interaction of this asymmetric dense shell with the CSM drives a reverse shock into the dense shell and outer layers of the ejecta and a forward shock into the surrounding CSM. The recombination of the shocked material produces \ion{H}{$\alpha$} emission at distinct velocities. The brighter emission component of \ion{H}{$\alpha$} at lower velocities ($\sim \pm 1950\, \mathrm{km\,s^{-1}}$) corresponds to the higher interaction power due to a more massive shell, while the weaker emission at higher velocities ($\sim \pm 5650\, \mathrm{km\,s^{-1}}$) arises from the less massive, faster-moving shell. This picture seamlessly aligns with the presence of two pairs of satellite components observed on either side of the central \ion{H}{$\alpha$} feature powered by radioactive decay (see Figure~\ref{fig:Ha_fit}). Moreover, unlike the models, we note that the high velocity \ion{H}{$\alpha$} components in SN~2023ixf display extended, slanted wings rather than sharp, vertical edges. \cite{Dessart2023} suggested that such wings can be explained by a less dense shell than assumed in the models, allowing some shock power to be reprocessed deeper in the ejecta, thus enhancing line emissivity over a broader velocity range.

The +259 d spectrum is also shown in the inset of Figure~\ref{fig:Comp_Dessart} to compare the extent of these features in the velocity space at +259 d and +363 d. While the high-velocity feature on the blue side of \ion{H}{$\alpha$} is distinct in the +259 d spectrum, an emerging high-velocity feature on the red side is also visible at this epoch. Since the high-velocity feature on the blue side blends with [\ion{O}{i}], it is challenging to determine at which wavelength it merges with the continuum. However, on the red side, it appears to merge with the continuum at a velocity of 9000$\,\mathrm{km\,s^{-1}}$. For the low-velocity components, blending with the central \ion{H}{$\alpha$} feature on one side and the high-velocity components on the other complicates the determination of their extent, but the multi-component fit indicates merging with the continuum at $\pm$5500$\,\mathrm{km\,s^{-1}}$. 

In the model spectra of \citet{Dessart2023} at various epochs (350, 500, and 700 d), features originating from the dense shell remain constrained within $\pm$8000$\,\mathrm{km\,s^{-1}}$, the velocity of injection of shock power at 350 d, while the strength of the feature increases over time. Therefore, for SN~2023ixf we can assume that the asymmetric dense shell expanding at velocities 5500 and 9000$\,\mathrm{km\,s^{-1}}$ (which are the respective merging points with continuum) along the equator and the poles respectively, is interacting with the extended CSM. The corresponding CSM radii, based on the distance travelled by the shock front over 259 days, are approximately 1.2 $\times$ 10$^{16}$ cm and 2.0 $\times$ 10$^{16}$ cm. Assuming the CSM expands at a typical red supergiant wind velocity of $10\,\mathrm{km\,s^{-1}}$, this CSM likely formed $\sim 390 - 640$ years prior to the explosion. However, it is important to note that these features may have emerged anywhere between +141 d to +259 d, this timeframe is approximate. Finally, a comparison with \citet{Dessart2023} models indicates that the minimum shock power required to produce the high-velocity feature is around 5$\times$10$^{40}$ erg s$^{-1}$.

Furthermore, we note that the \ion{H}{$\alpha$} profile shows asymmetry: the blue-shifted feature is stronger and exhibits a triangular shape, while the red-shifted feature is weaker and displays a boxy profile as shown in the inset of Figure~\ref{fig:Comp_Dessart}. Such red-blue asymmetry is also seen in the blueward components of [\ion{O}{i}] doublet and [\ion{Ca}{ii}] lines, and, to a lesser extent, in the \ion{Ca}{ii} NIR triplet. Dust formation likely contributes to this asymmetry, with redshifted radiation more attenuated as it traverses dusty regions. This asymmetry, first evident in the +125 d spectrum \citep{Singh2024}, likely arises from the formation of dust in a cold dense shell (CDS; \citealt{Chugai2009}) produced during the interaction of ejecta with CSM expelled within a decade prior to the explosion. As discussed above, the +363 d spectrum indicates interaction with a more distant CSM, which could lead to the formation of additional CDS, providing a secondary source of dust that enhances the observed red-blue asymmetry. The increasing prominence of this asymmetry over time strongly supports ongoing dust formation as SN~2023ixf evolves. This dust-driven asymmetry, coupled with CSM interaction with an aspherical dense shell, underscores the interplay of geometry and dust formation in shaping the observed spectral features.

\begin{figure}
	\includegraphics[width=0.53\textwidth, clip, trim={2cm, 0, 0, 3.0cm}]{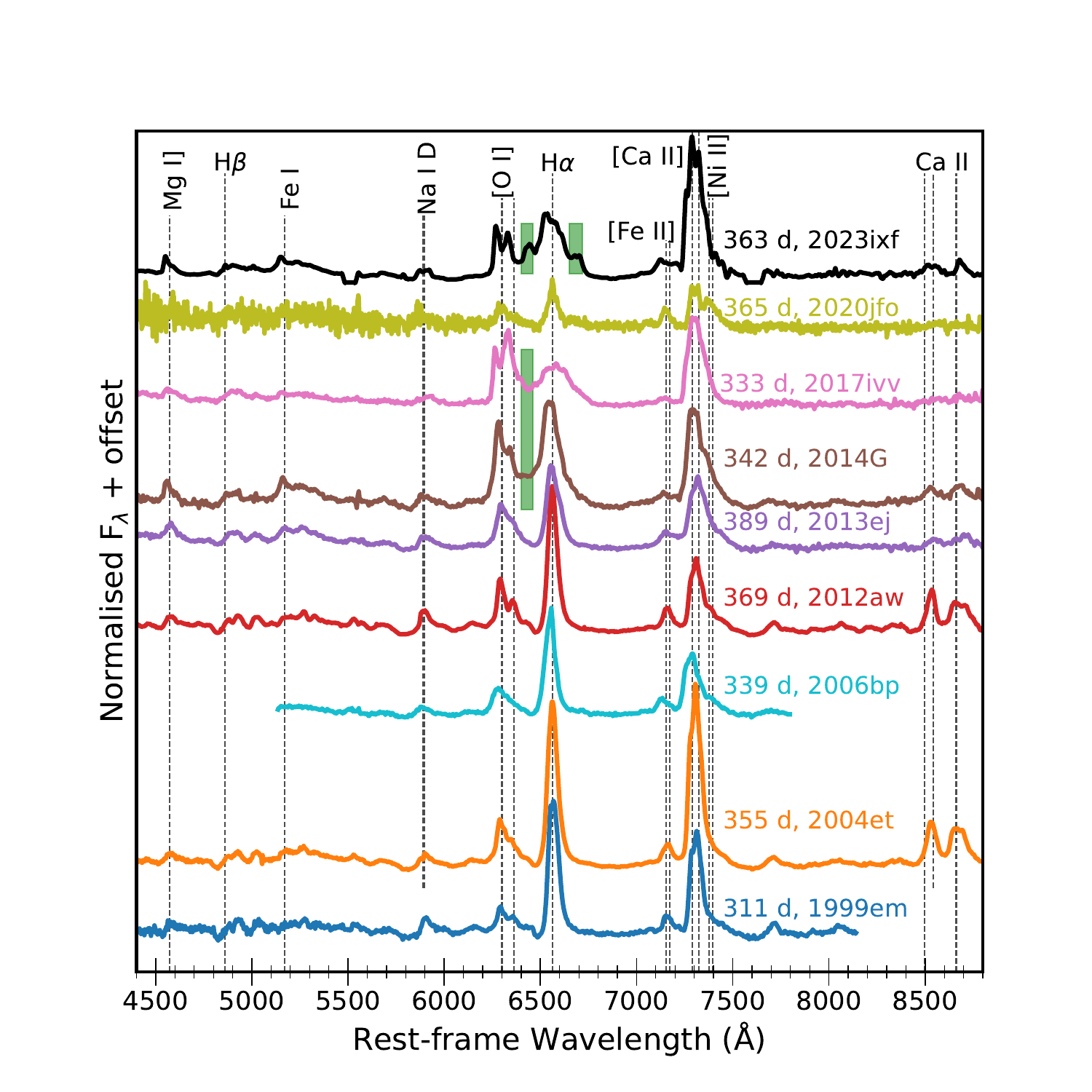}
    \caption{Comparison of the +363 d spectrum of SN~2023ixf with similar epoch spectra from the comparison sample. Interaction signatures in SNe~2014G, 2017ivv, and 2023ixf are highlighted with green bands.}
    \label{fig:spec_comp_sample}
\end{figure}

\begin{table}
\caption{Parameters of Type II SNe used in Figure~\ref{fig:spec_comp_sample} along with the references for the explosion epoch and/or spectra. The redshifts are taken from NED. \label{comp_sample}}
\centering
\renewcommand{\arraystretch}{1.1}
\setlength{\tabcolsep}{3pt}
\footnotesize
\begin{tabular}{llccc}
\hline
SN      & Explosion  & Redshift & Spectra Epoch & References \\
        & Epoch (MJD) &          &    (d)        \\
\hline
1999em  & 51474.7  & 0.002390        & 311 & [a] \\
2004et  & 53270.5  & 0.000909        & 355 & [b] \\
2006bp  & 53834.6  & 0.003502        & 339 & [c], [d] \\
2012aw  & 56002.6  & 0.002600        & 369 & [e], [f] \\
2013ej  & 56497.5  & 0.002190        & 389 & [g] \\
2014G   & 56669.3  & 0.004500        & 342 & [h] \\
2017ivv & 58091.6  & 0.03            & 333 & [i] \\
2020jfo & 58973.1  & 0.005220        & 365 & [j] \\
2023ixf & 60082.7  & 0.000804        & 363 & [k], This work   \\
\hline
\end{tabular}
\newline
{\footnotesize
References: [a] \cite{Leonard2002}, [b] \cite{Faran2014}, [c] \cite{Quimby2007}, [d] \cite{Inserra2012}, [e] \cite{Fraser2012}, [f] \cite{Jerkstrand2014}, [g] \cite{Yuan2016}, [h] \cite{Terreran2016},  [i] \cite{Gutierrez2020}, [j] \cite{Ailawadhi2023},  [k] \cite{Hsu2024} and references therein.}
 \end{table}

\section{Comparison with other Type II SNe}\label{sec:spec_comp_sample}

We compare the +363 d spectrum of SN~2023ixf with those of other Type II SNe that have spectra available within $\pm$50 days of this epoch. From the literature, we obtained spectra for the following SNe: 1999em, 2004et, 2006bp, 2012aw, 2013ej, 2014G, 2017ivv and 2020jfo. The redshifts and explosion epochs, and the corresponding references for these SNe, as considered in this work, are provided in Table~\ref{comp_sample}. SNe~1999em, 2004et, 2006bp, 2012aw, 2013ej and 2020jfo do not show broad boxy components of \ion{H}{$\alpha$}, considered signs of late-time interaction. Interaction signatures are evident in SNe~2014G and 2017ivv, though they are not as pronounced as those in SN~2023ixf. This indicates the presence of a dense shell in SN~2023ixf, formed earlier than in SNe~2014G and 2017ivv, and hence located closer to the progenitor. The distance of this CSM shell depends on the progenitor’s wind mass-loss rate, the wind velocity, and the density and velocity of the ambient medium \citep{Kurfurst2020}. SN~2014G displays a bridge-like feature connecting [\ion{O}{i}] and \ion{H}{$\alpha$}, caused by broad \ion{H}{$\alpha$} emission — a sign of interaction — which is also seen in SN~2023ixf but over a shorter wavelength range. SN~2017ivv also exhibits a broad \ion{H}{$\alpha$} component that enhances the red component of the [\ion{O}{i}] doublet profile \citep{Dessart2022}. The blue-shifted [\ion{O}{i}] feature observed in SN~2023ixf is also apparent in SN~2017ivv. None of the comparison SNe, except SN~2023ixf, exhibit the red-blue asymmetric peaks in \ion{H}{$\alpha$} and [\ion{Ca}{ii}], which are considered signatures of dust formation. These observations highlight SN~2023ixf as a uniquely interacting and dynamically evolving Type II SN, showcasing signatures of a dense, close-in CSM shell and potential dust formation that distinguish it from the other SNe in the comparison sample.

\section{Discussion and Conclusion}\label{sec:conclusion}

In this study, we analysed the nebular phase spectrum of SN~2023ixf obtained at one-year post-explosion using the recently commissioned WEAVE multi-object fibre-fed spectrograph at the prime focus of 4.2m WHT. This spectrum of SN~2023ixf, the closest Type II SN of the decade, represents a significant landmark as it marks the first SN spectrum ever captured with WEAVE. This milestone highlights the instrument's capabilities for detailed spectroscopic analysis of transient phenomena, particularly in the late phases of SN evolution.

To investigate the spectral evolution of SN~2023ixf, we incorporated published nebular spectra from earlier epochs (+141 d and +259 d). The evolution of emission features over time reveals a rich array of lines typically observed in Type II SNe. Additionally, narrow host galaxy emission lines, such as \ion{H}{$\beta$} and [\ion{O}{iii}] $\lambda$5007, are detected at nearly zero velocity. 

The +363 d spectrum exhibits a complex \ion{H}{$\alpha$} profile, a hallmark of ongoing interaction between the SN ejecta and CSM. Detailed multi-Gaussian decomposition of this profile reveals that the central broad feature spanning $\pm$4000$\,\mathrm{km\,s^{-1}}$ can be decomposed into three Gaussian components: a red-shifted feature near 12\,$\mathrm{km\,s^{-1}}$ and two symmetric components centred at $\pm$1950$\,\mathrm{km\,s^{-1}}$. Furthermore, two high velocity components centred at $\pm$5650$\,\mathrm{km\,s^{-1}}$ emerges in the +363 d spectrum. The two pairs of satellite components indicate interactions of an asymmetric shock-swept dense shell with the surrounding CSM expelled a few hundred years before the explosion.

Studies of emission lines (mainly [\ion{O}{i}] doublet and \ion{Mg}{i}]) in the nebular-phase spectrum of SN~2023ixf at +363 d highlight that the geometry of SN~2023ixf’s ejecta does not require a toroidal or disk-like structure to explain the observed double-peaked [\ion{O}{i}] profile. Instead, the symmetric peaks arise naturally from the intrinsic doublet nature of [\ion{O}{i}] $\lambda\lambda$6300, 6364. The observed blue shift of both [\ion{O}{i}] and \ion{Mg}{i}] suggests the presence of a moderately blueshifted clump or shell of oxygen-rich material moving at $\sim$1500 km s$^{-1}$ toward the observer, overlaid on a broader, near-zero velocity component that likely represents a more spherically symmetric distribution of ejecta. The lower-than-expected [\ion{O}{i}] doublet ratio ($\sim$1.15) further implies optically thick line emission. These findings indicate a complex ejecta structure with a central symmetric distribution and an additional, directional component rather than a purely toroidal or disk-like morphology.

The analysis of the [\ion{O}{i}] doublet flux, combined with comparisons of the +363 d spectrum to modelled spectra, suggests $M_{\rm O} \approx 0.07-0.30\,M_\odot$, $M_{\rm He}$ $\lesssim$0.3 $M_\odot$ and $M_{\rm ZAMS}$ $\lesssim$12 $M_\odot$ for SN~2023ixf. The progenitor mass of SN~2023ixf estimated in this study aligns with numerous previous studies that suggested a low-mass progenitor for SN~2023ixf, based on pre-explosion imaging and hydrodynamical modelling. 

The comparison of SN~2023ixf +363 d spectrum with other Type II SNe around the same epoch shows that while most (e.g., SNe 1999em, 2004et, 2006bp, 2012aw, 2013ej, 2020jfo) lack significant late-time interaction signatures, SNe 2014G and 2017ivv show moderate interaction features. In contrast, SN~2023ixf exhibits stronger interaction signatures, indicating a close-in CSM shell formed earlier and located closer to the progenitor. Additionally, SN~2023ixf uniquely displays red-blue asymmetric peaks in \ion{H}{$\alpha$} and [\ion{Ca}{ii}], likely due to dust formation, setting it apart from the comparison sample. This highlights the distinct evolution and interaction dynamics of SN~2023ixf in comparison to other Type II SNe.

This study underscores the importance of further investigation of SN~2023ixf, focusing on the $UV$ and the evolution of the \ion{H}{$\alpha$} profile, particularly its blueward and redward components. These regions prominently exhibit the effects of the shock generated by the ejecta-CSM interaction and offer significant potential to constrain the interaction properties and shock power as a key energy source at late times. Overall, continued late-time follow-up of SN~2023ixf will provide valuable insights into the mass-loss history of its progenitor star, offering a deeper understanding of its evolution thousands of years before core collapse.

\section*{Data Availability}
Data can be shared upon request to the corresponding author.

\section*{Acknowledgement} 

AK and JRM are supported by the UK Science and Technology Facilities Council (STFC) Consolidated grant ST/V000853/1. RD acknowledges funds by ANID grant FONDECYT Postdoctorado Nº 3220449.

The authors thank Javier Mendez Alvarez for their support in preparing the WEAVE observations. The authors further express their gratitude to Luc Dessart for providing the modelled spectra published in \cite{Dessart2023} and Avinash Singh and Lucia Ferrari for sharing the +141 d and +259 d spectra, respectively.

Based on observations made with the William Herschel Telescope operated on the island of La Palma by the Isaac Newton Group of Telescopes in the Spanish Observatorio del Roque de los Muchachos of the Instituto de Astrofísica de Canarias (WEAVE proposal WS2024A2-005).

Funding for the WEAVE facility has been provided by UKRI STFC, the University of Oxford, NOVA, NWO, Instituto de Astrofísica de Canarias (IAC), the Isaac Newton Group partners (STFC, NWO, and Spain, led by the IAC), INAF, CNRS-INSU, the Observatoire de Paris, R\'{e}gion \^{I}le-de-France, CONACYT through INAOE, the Ministry of Education, Science and Sports of the Republic of Lithuania, Konkoly Observatory (CSFK), Max-Planck-Institut f\"{u}r Astronomie (MPIA Heidelberg), Lund University, the Leibniz Institute for Astrophysics Potsdam (AIP), the Swedish Research Council, the European Commission, and the University of Pennsylvania.  The WEAVE Survey Consortium consists of the ING, its three partners, represented by UKRI STFC, NWO, and the IAC, NOVA, INAF, GEPI, INAOE, Vilnius University, FTMC – Center for Physical Sciences and Technology (Vilnius), and individual WEAVE Participants. Please see the relevant footnotes for the WEAVE website\footnote{\url{https://weave-project.atlassian.net/wiki/display/WEAVE}} and for the full list of granting agencies and grants supporting WEAVE\footnote{\url{https://weave-project.atlassian.net/wiki/display/WEAVE/WEAVE+Acknowledgements}}.

\bibliographystyle{mnras}
\bibliography{manu}

\bsp
\label{lastpage}
\end{document}